\documentclass[11pt]{article}

\usepackage[margin=1.15in]{geometry}
\usepackage{graphicx}
\usepackage{booktabs}
\usepackage{amsmath}
\usepackage{xcolor}
\usepackage[round]{natbib}
\usepackage{microtype}
\usepackage[colorlinks=true,linkcolor=blue!55!black,citecolor=blue!55!black,urlcolor=blue!55!black]{hyperref}

\graphicspath{{figures/}}

\newcommand{\OR}[1]{OR~#1}

\title{Invisible to the Machine: Auditing AI Restaurant,\\ Caf\'e, and Bar Recommendation Against a\\ Complete Market Census}

\author{Vladimir Pitenin\thanks{Norly Research (\url{https://norly.co}). This study was funded and conducted by Norly; Section~\ref{sec:ethics} discloses the competing interest and the insulation measures. Correspondence: \texttt{dino@norly.co}.}\\ Norly Research}

\date{August 2026}

\begin{document}

\maketitle

\begin{abstract}
\noindent AI assistants are becoming a primary interface for local discovery, yet almost nothing is known about which venues they surface --- especially in food and drink, where recommendations carry direct revenue consequences. We present the first census-denominated audit of AI venue recommendation: a complete enumeration of 4,776 caf\'es, restaurants, and bars across two bounded markets (Canggu and Ubud, Bali), against which we evaluate 2,208 search-grounded responses from four production AI systems (ChatGPT, Claude, Gemini, Perplexity) to 96 persona-conditioned queries, collected over seven days under a pre-registered protocol. Because we observe the full market, we can measure what sampled audits cannot: 85.6\% of venues were never recommended by any system --- 72.6\% even among established venues with fifty or more ratings. Visibility follows a two-margin structure. Entry into answers is associated with documentation: review volume (\OR{1.64}), an own website (\OR{1.92}), listed price information (\OR{1.54}), and third-party web mentions (\OR{1.44}) --- while star rating is null at this margin (\OR{0.89}). Rank within answers reverses the pattern: among recommended venues, rating significantly predicts first position (\OR{1.17}). Presence in an open POI dataset (Foursquare), a folk-theorized visibility factor, shows no positive effect at either margin. Outright fabrication is rare (0.08\% of mentions), but systems recommended permanently closed venues 93 times --- staleness, not hallucination, is the practical failure mode. Cross-system agreement is low (top-20 Jaccard 0.33--0.54). A two-week test--retest shows cross-period answer similarity comparable to same-day rerun similarity: the churn is sampling stochasticity, not temporal drift. We release our protocol, registry construction method, and derived data.
\end{abstract}

% =====================================================================
\section{Introduction}
\label{sec:intro}

A growing share of local discovery now runs through conversational AI. When a traveler asks an assistant where to work over coffee in Canggu or where to take a date in Ubud, the answer arrives as a short, synthesized list --- three to eight venues, plucked from a market of thousands, with no page two. For the caf\'es, restaurants, and bars on the wrong side of that synthesis, the consequences are not hypothetical: two decades of evidence show that how a venue is represented on discovery intermediaries carries causally identified demand effects --- an extra Yelp star raises independent-restaurant revenue by 5--9\% \citep{luca2016reviews}, and crossing a displayed half-star threshold makes evening sellouts 19 percentage points more likely \citep{anderson2012learning}. The chain now extends into the AI channel itself: an assistant's recommendation lifts same-brand searches by 4.3 percentage points among previously unengaged consumers \citep{iannelli2026prompt}. Yet these new systems do not merely rank venues --- they speak about a chosen few and omit the rest entirely, and what they actually surface for food and drink, and what separates the surfaced from the invisible, has no scientific measurement (vendor benchmarks of single systems exist and are discussed in Section~\ref{sec:related}).

The nascent audit literature approaches the question from two directions, each with a structural blind spot. Conjoint experiments randomize venue attributes within constructed choice sets --- the strongest causal design available --- but by construction place every candidate before the model, so they cannot observe whether a real venue enters consideration at all; the closest such study audits hotels \citep{baig2026hotel}. Large-scale prompt audits measure which brands surface across thousands of runs, but they evaluate against curated catalogues of tens to hundreds of prominent brands \citep{zatuchin2026owns, jack2026prominence}, so they can report \emph{relative} prominence but not a population rate: absent a denominator, ``never recommended'' is unmeasurable. And no study in either tradition has examined food and drink venues --- the segment where independent operators dominate, documentation is thinnest, and the revenue evidence above is strongest.

We close both gaps with a census. We enumerate the complete food-and-drink market of two bounded submarkets --- 4,776 operating caf\'es, restaurants, and bars across greater Canggu and greater Ubud, Bali --- and audit what four production AI systems (ChatGPT, Claude, Gemini, Perplexity), queried through their search-grounded interfaces, recommend across 96 persona-conditioned queries repeated over seven days: 2,208 runs, 12,439 valid venue mentions, and 1.4~million venue-level exposure opportunities in the primary analysis frame, collected under a protocol pre-registered before confirmatory collection began. Because every mention is resolved against the full census --- with extraction and entity matching validated by independent double annotation and audited against measured error rates --- we can compute what sampled audits cannot: the population rate of AI invisibility, and the venue-level factors associated with escaping it.

Five findings organize the paper. First, invisibility is the norm: at least 85.6\% of venues were never recommended by any system in any run --- 72.6\% even among established venues with fifty or more ratings --- and these are floors: under the broadest defensible market frame the rate exceeds 92\% (Appendix~\ref{app:census}). Yet visibility among the surfaced is long-tailed rather than winner-take-all (the leading venue holds 1.9\% of recommendations). Second, visibility is two-margined --- documentation admits, rating ranks: entry into answers is associated with review volume (\OR{1.64}), an own website (\OR{1.92}), listed price data (\OR{1.54}), and web mentions (\OR{1.44}), while star rating is null at entry (\OR{0.89}) and emerges only in ordering venues within answers (\OR{1.17} for first position). This dissociation reconciles the conjoint and audit traditions: they measure different margins of one pipeline. Third, a widely assumed factor fails its first direct test: presence and standing in an open point-of-interest dataset (Foursquare) predicts nothing positive at either margin once documentation is controlled. Fourth, the systems' practical failure mode is staleness, not fabrication: one likely-invented venue name in more than 12,000 valid mentions, against 93 recommendations of permanently closed venues. Fifth, answers are unstable run-to-run and across paraphrases, yet a two-week test--retest shows no temporal drift beyond that noise floor --- visibility is a persistent venue property measured through a stochastic channel, and single-shot visibility checks measure noise.

Beyond these results, the study contributes methodologically. We show that audit conclusions are sensitive to instrument quality in ways this literature has not examined: a matching-error remediation reversed one factor's apparent effect from significantly positive to null, and one covariate was contaminated by our own collection path --- both caught by pre-registered validation gates and reported with measured error rates. We release the census-construction method, protocol, and derived data for replication in other markets.

The paper proceeds as follows: Section~\ref{sec:related} situates the work; Section~\ref{sec:context} describes the market census and collection apparatus; Section~\ref{sec:method} details extraction, matching, validation, and the pre-registered models; Section~\ref{sec:results} reports results; Section~\ref{sec:discussion} discusses mechanisms, implications for venues, platforms, and the emerging GEO industry, and the methodological lessons; Section~\ref{sec:limitations} states limitations; Section~\ref{sec:ethics} covers ethics and disclosure, including the funder's commercial interest and the measures taken to insulate the findings from it.

% =====================================================================
\section{Related Work}
\label{sec:related}

\subsection{LLMs as recommenders and their biases}

Large language models increasingly function as de facto recommenders, and a benchmark literature shows they do so unevenly: recommendation output is swayed by candidate order and item popularity \citep{hou2024llmrankers} --- a caution for designs that supply candidate lists. Our audit elicits open-ended recommendations instead, a setting where position effects over supplied candidates cannot arise but popularity effects can, and do. Persona-conditioned prompting shows the same instability along social lines: lists shift with attributes like race, gender, or nationality \citep{zhang2023chatgpt}, and a parallel brand-level bias favors global and chain brands over local ones \citep{kamruzzaman2024global}, though the direction is contested --- one study finds an LLM recommender \emph{less} popularity-biased than classical collaborative filtering \citep{lichtenberg2024popularity}. The cold-start setting --- no history, no declared attributes --- reproduces consistent cultural stereotyping \citep{andre2025biases} and is the representative condition for local discovery. \citet{ma2026fairness} survey the resulting fairness metrics and motivate the user-side/item-side distinction we borrow.

\subsection{Generative engine optimization and visibility audits}

A second, more applied literature asks not whether LLM recommendations are biased but whether visibility in generative answers can be measured and moved --- generative engine optimization (GEO). \citet{aggarwal2024geo} opened the question: content-side edits (citations, statistics, authoritative phrasing) can raise a source's visibility by up to roughly 40\%, an effect \citet{kumar2024manipulating} pushed to its adversarial limit, forcing a rarely-recommended product into an LLM's top slot via optimized text sequences. \citet{martinez2026optimizing} supplies the field's critical synthesis --- GEO as a multi-stage, partially observable pipeline --- showing the widely cited $\sim$40\% figure holds only if a source already occupies the retrieved context. We adopt this survey's reproducibility bar --- repeated measurement, paraphrase controls, human-validated extraction --- as our standard.

The closest published precedent is \citet{baig2026hotel}: a pre-registered conjoint randomizing guest rating, price, and other attributes across twelve LLMs choosing among hotels, finding a top rating raises selection probability 31.6 percentage points and a high price lowers it 30.0 points. Two things bound our contribution here. \citeauthor{baig2026hotel} study hotels; food and drink, where independent operators rather than chains are the modal case, remains untouched by attribute-level audit. More fundamentally, their conjoint is experimental and ours observational: randomizing attributes within a fixed choice set licenses a causal claim about what happens once a venue is already competing for attention --- a claim our design does not attempt. Enumerating a complete market instead of constructing a choice set lets us ask whether a venue enters consideration at all.

That census denominator also separates us from the two audits closest to us in scale. Running roughly 37,000 queries against a 533-brand catalogue, \citet{jack2026prominence} find long-tail brands facing ``catastrophic invisibility'' --- 48--52\% never surface at all; running 3,750 queries across 50 brands, \citet{zatuchin2026owns} finds only moderate concentration (Gini 0.28) and low cross-model agreement on the top brand (41.6\%) --- evidence against a winner-take-all reading of AI recommendation. Neither catalogue enumerates a real market, so neither yields a population-level invisibility rate; our registry of 4,776 venues across two Bali submarkets is the denominator that lets us report one. Where comparable, results converge: our cross-engine top-20 Jaccard (0.33--0.54) echoes \citeauthor{zatuchin2026owns}'s low agreement in direction if not magnitude. \citet{jack2026paraphrase} add a warning our design tests rather than assumes: paraphrase variance exceeds rerun variance --- a pattern now confirmed for production Google, where AI Overviews front 51.5\% of real-user queries yet are less consistent across reruns and less robust to minor edits than the organic results they displace \citep{grossman2026disrupts}, and longitudinal brand tracking at the 100,000-prompt scale documents scale-dependent visibility disparities disadvantaging smaller entities \citep{kumar2026geoscale}. \citet{chen2025dominate} document a systematic sourcing bias of AI search toward earned third-party media over brand-owned content --- a supply-side pattern our venue-level citation data partially complicates (Section~\ref{sec:sources}). Closest to our vertical, commercial vendors of visibility-tracking tools have recently circulated non-peer-reviewed benchmarks reporting that large majorities of restaurants (figures in the 75--83\% range) never appear in single production systems' AI answers. These analyses are self-interested, cover one system each, and lack factor models, validated extraction, and market denominators; we note them as directional context only --- the phenomenon they gesture at is what our census-denominated, multi-system design measures under a disclosed and validated protocol. A peer-reviewed miniature of the same phenomenon: prompted repeatedly for German Christmas markets, a production model reproduces a narrow canon of roughly five icons from a real pool of $\sim$2,000, under hidden selection criteria \citep{spennemann2026travel} --- the canon effect our concentration analysis quantifies at market scale.

Two further papers shape our outcome variable. \citet{zhang2026citation} separate citation, absorption, and influence as distinct constructs rather than one ``visibility'' measure; we adopt the distinction, coding mentions as recommended, mentioned-neutral, or advised-against. \citet{seo2026verified} document that search-augmented systems routinely cite sources that do not support the claim made --- confidently wrong but ``sourced'' --- the antecedent of our unmatched-mention taxonomy, which our census lets us build with an unusual guarantee: we can show a recommended name has no living referent. Finally, nothing in this literature tests whether presence in an open, third-party point-of-interest dataset predicts AI visibility, despite being a common industry assumption. Our Foursquare test is, to our knowledge, the first of its kind; the answer, once other signals are controlled, is a plain null.

\subsection{Local discovery and hospitality}

Bias in algorithmically mediated local discovery predates generative AI: ``near me'' retrieval already concentrates exposure unevenly across nearby businesses \citep{banerjee2020nearme}, a lineage traced to \citet{introna2000shaping}'s argument that search intermediaries systematically privilege already-prominent content. In the LLM era, reissuing identical restaurant requests across English dialects and code-switched variants shifts which restaurants get recommended \citep{venkateswaran2026linguistic} --- aimed at linguistic rather than venue-level bias, leaving open which venue characteristics actually drive food-and-drink recommendation. Socioeconomic geography matters too: LLM travel recommendations for lower-income countries are less unique and reference fewer specific locations \citep{bhagat2024richer}, a caution given our single-region, English-only design. Two hospitality-journal reviews confirm the gap: both find the field studies consumer adoption of tools like ChatGPT, not what those tools output \citep{zhang2026hospitality, tuyen2025chatgpt} --- whether travelers \emph{use} generative AI, not what it \emph{tells them}.

The economic stakes of that gap are established for the pre-LLM case. Exploiting Yelp's rounding of ratings to a displayed half-star, \citet{luca2016reviews} identifies a causal effect of rating on revenue: a one-star increase raises restaurant revenue 5--9\%, concentrated entirely in independent restaurants. Applying the same logic to reservation availability, \citet{anderson2012learning} find that crossing a displayed half-star threshold makes a restaurant 19 percentage points --- roughly 49\% in relative terms --- more likely to sell out its evening seating. Both studies show that how a restaurant is represented to a discovery intermediary carries causally identified demand consequences before any AI system entered the pipeline. The chain now extends into the AI channel itself: joining clickstream panels to the same users' assistant conversations, \citet{iannelli2026prompt} show that an AI recommendation to a previously unengaged consumer lifts same-brand searches by 4.3 percentage points and brand-site visits by 2.4 --- AI recommendations demonstrably move consumers, which makes who gets recommended an economic question rather than a curiosity. \citet{li2021yelp} supply the direct pre-LLM precedent for our design: the same restaurants receive different ratings on Yelp versus Google Maps, and each platform's top-ten list for a metro barely overlaps the other's.

\subsection{Audit methodology}

We situate our data-collection bot within the algorithm-audit tradition: \citet{sandvig2014auditing} name and legitimize exactly this design --- a noninvasive, sock-puppet audit that programmatically impersonates users and records platform responses --- as one of the few methods available when a platform's ranking logic is not otherwise inspectable. \citet{bouchaud2024auditing} find this genre chronically weakened by undisclosed, arbitrary design choices that platform opacity forces on auditors, rarely justified in print. We respond directly: our research questions, hypotheses, and analysis plan were pre-registered before confirmatory collection began.

We also extend the tradition with a result it has not previously reported: factor-importance conclusions in this kind of audit are sensitive to entity-matching quality. A remediation pass fixing name-matching errors, with no change to the underlying data, reversed our preliminary finding for third-party data-provider presence from a significant positive association (\OR{1.70}) to a null effect (\OR{0.84}) in the confirmatory model, because the original matcher's errors correlated with data-provider absence. No prior audit we know of validates its matching pipeline this way.

Taken together, the conjoint evidence that rating dominates hotel selection and the concentration evidence that cross-model agreement is low are reconcilable with our results once the margin being measured is made explicit. The conjoint of \citet{baig2026hotel} necessarily measures a within-choice-set margin: every hotel profile is, by construction, already before the model, so their rating effect describes what happens once a venue is being considered. Our census lets us measure that margin and an earlier one. Restricted to venues an engine actually recommends, rating significantly predicts first position in our data (\OR{1.17}) --- consistent with \citeauthor{baig2026hotel} at the only margin their design can observe. But at the earlier margin their conjoint cannot pose --- whether a venue enters an answer at all --- rating is null (\OR{0.89}) once review volume is controlled; entry instead tracks documentation. The two studies describe one two-margin structure, not a contradiction: reputation governs rank once a venue is in the running; discoverability infrastructure governs whether it gets there at all.

% =====================================================================
\section{Study Context and Data}
\label{sec:context}

\subsection{Setting}
\label{sec:setting}

We study two adjacent submarkets on Bali, Indonesia: greater Canggu (including Berawa, Batu Bolong, and Pererenan) and greater Ubud (including Penestanan, Sayan, and Pengosekan), delimited by fixed geographic polygons (coordinates in the released configuration). The setting offers three properties an audit of AI venue recommendation needs: a bounded, walkable market small enough to enumerate completely; an English-language query population --- tourists and resident remote workers --- for whom conversational AI is a natural discovery channel; and an extremely dense, competitive food-and-drink sector dominated by independent operators, the segment for which the revenue stakes of discovery are best documented (Section~\ref{sec:related}) and the audit literature is silent.

\subsection{The venue census}
\label{sec:censusframe}

Our population frame is defined operationally: \textbf{all places listed on Google Places under five food-service types (restaurant, cafe, bar, coffee\_shop, bakery) within the study polygons at census time, augmented by resolution of AI-recommended names.} Google Places is simultaneously the dominant consumer discovery substrate and a principal grounding source for the audited systems, making it the appropriate frame for the population AI systems could plausibly surface.

Enumeration used the Places Nearby Search API over an adaptive grid. Because the API returns at most 20 results per query, any saturated cell (returning exactly 20) was recursively subdivided into four half-radius cells down to a 130~m floor --- without this, dense centers silently truncate and the ``census'' undercounts precisely where venues concentrate. The completed grid required 760 requests across the two polygons. The frame was then augmented by a resolution pass: names recommended by the AI systems that failed to match any census venue were probed against Places Text Search (with locality and bounding guards), adding venues of atypical listing type --- hotel restaurants, coworking caf\'es --- that food-type enumeration misses. Roughly 220 candidate names were probed across two passes; \textbf{70 venues in the final census carry this lookup provenance} (most probes resolved to venues the grid census already held --- itself evidence of frame adequacy, quantified in Appendix~\ref{app:census}). The final census contains \textbf{4,776 venues}, each carrying its Google profile snapshot (rating, review count, price level, hours, website, business status) at collection time.

No census of a living market is complete in an absolute sense. Appendix~\ref{app:census} reports a dual-frame capture--recapture analysis against an independent enumeration (the Foursquare open POI dataset), a hand audit of the non-overlapping stratum, and an adversarial-probe analysis showing that the systems' own 16,447 mentions (pilot, wave, and retest) resolve overwhelmingly to the census. Because any venue absent from the frame is, by that analysis, almost surely never recommended, all reported invisibility rates are floors (Section~\ref{sec:invisible}).

\subsection{Systems under audit}
\label{sec:systems}

We audit four production assistants through their public, search-grounded APIs, the access route used by prior audits \citep{jack2026prominence, martinez2026optimizing}: OpenAI (gpt-5.2-2025-12-11, Responses API with the web-search tool), Anthropic (claude-sonnet-5 with the server-side web-search tool), Google (gemini-3.5-flash with Google Search grounding), and Perplexity (sonar). Configurations approximate each vendor's consumer default tier; model version and timestamp are logged on every run. Two configuration notes: Claude's search budget was capped at two searches per run (a disclosed cost decision made before the confirmatory wave; the cap bound in practice --- 97.6\% of the Claude arm's wave runs consumed both searches --- and a sensitivity refit drops the Claude arm, Section~\ref{sec:entry}); Gemini's grounding API exposes no user-location parameter, so location context for all systems is carried in the query text itself, with approximate location hints additionally supplied where the API supports them.

\subsection{Query instrument}
\label{sec:instrument}

Queries simulate realistic, persona-conditioned local discovery: eight personas (digital nomad seeking a work caf\'e, couple planning a date, business meeting host, budget backpacker, family with children, vegan/dietary-constrained diner, specialty-coffee enthusiast, late-night group) $\times$ six first-person narrative templates per persona $\times$ two areas $=$ \textbf{96 unique queries}. Templates are deliberate paraphrases of a constant intent (example: ``I'm 35, just arrived in Canggu, Bali and I'll be staying for about a month. I work remotely and I'm looking for a cafe where I can work on my laptop for a few hours --- good wifi, decent coffee, and it shouldn't be too loud. Where should I go?''), enabling the paraphrase-sensitivity analysis of Section~\ref{sec:stability}. The instrument was designed by the research team at the start of the project, anchored on a real traveler-style query, and frozen at pre-registration; the pre-registration was published before confirmatory collection began. The complete instrument is reproduced verbatim in Appendix~\ref{app:queries} and released in machine-readable form with the replication package.

\subsection{Collection design}
\label{sec:collection}

The confirmatory wave ran over seven calendar days, with repetitions deliberately spread across days: Perplexity 10 repetitions per query, OpenAI and Gemini 5, Claude 3 --- unequal by budget design, handled by the estimation strategy (Section~\ref{sec:models}) --- for \textbf{2,208 runs}, where a run is one rendered query submitted once to one system, returning one search-grounded answer. OpenAI and Claude arms ran through the vendors' asynchronous batch APIs; Gemini and Perplexity ran synchronously with pacing. A pre-registered test--retest holdout (16 queries $\times$ 4 engines, 144 runs) was collected two weeks after the wave (Appendix~\ref{app:retest}). An exploratory pilot (544 runs) preceded pre-registration and informed instrument design only; no confirmatory statistic uses pilot data.

% =====================================================================
\section{Method}
\label{sec:method}

\subsection{Mention extraction and validation}
\label{sec:extraction}

Each response is parsed by an LLM extractor (Claude Haiku with a strict JSON schema and a schema-free fallback prompt) into venue mentions carrying the name as written, rank of appearance, and sentiment (recommended / mentioned-neutral / advised-against --- adopting the construct distinction of \citealp{zhang2026citation}). A deterministic post-extraction filter, with blocklists committed to the repository, removes classes the written extraction rules exclude: platforms named as venues, coworking and convenience-store brands, bare generic categories, and within-run duplicates (259 wave mentions; 2.0\% of raw wave extractions). The wave yields \textbf{12,439 valid mentions}.

Extraction was validated by independent double annotation: a second, separately prompted model labeled a stratified 95-run sample blind to the extractor's output, and a human adjudicator resolved all 40 disagreement runs against the written rules (one sampled run was excluded after the annotator disclosed an independence breach during setup --- its procedure, not its content, was disqualifying). Adjudicated performance: \textbf{mention-level precision 97.6\%, recall 99.1\%; strict run-level agreement 84.0\%}, rising to \textbf{91.5\%} after the class-based remediation above was applied and re-scored against the frozen labels. Residual error classes (five low-salience misses; one entity confusion --- a coffee variety extracted as a venue) are reported rather than patched: the remediation blocklists target error \emph{classes}, not individual names found in validation, so the re-scored gate cannot pass by construction. Against the pre-registered $\geq$95\% accuracy gate, whose metric level was left unspecified, we report all three numbers and note the mention-level metrics --- the standard for information extraction --- clear it while strict run-level does not.

\subsection{Entity matching and its audit}
\label{sec:matchingsec}

Mentions are resolved to census venues by a deterministic fuzzy matcher operating on distinctive-token cores (generic and geographic tokens stripped): full-name similarity $\geq$87 with a core-similarity veto $\geq$80; containment acceptance (mention core fully contained in a candidate core); core similarity $\geq$90 alone; and a cross-area guard requiring multi-token containment or near-exact similarity before crediting a venue in the other study area, with same-area candidates winning ties. A final, evidence-audited pass applies 142 name-variant mappings verified during the unmatched-mention taxonomy (Section~\ref{sec:predictors}) --- rebrands, nicknames, accent variants --- each recorded under a distinct match-method flag. \textbf{87.9\% of valid wave mentions resolve to the census.}

Matching was audited on a 100-mention sample deliberately over-weighting low-score strata: population-weighted correct-entity accuracy is \textbf{$\approx$98--99\%}, with errors concentrated below score 95 (near-name pairs, ambiguous short brand names); a sensitivity refit excluding sub-95 matches (Section~\ref{sec:entry}) tests conclusions against this stratum. During development, validation evidence forced one matcher revision whose consequences we report in full: the revision reversed a predictor's apparent effect (Section~\ref{sec:lessons}) --- the reason audit pipelines need instrument validation, not just outcome reporting.

\subsection{Outcome construction}
\label{sec:outcome}

The unit of analysis is venue $\times$ persona $\times$ engine. For venue $v$, persona $p$, engine $e$: trials $=$ wave runs of same-area queries of $p$ on $e$ (6 templates $\times$ the engine's repetitions); successes $=$ trials whose response carries a valid, matched, recommended mention of $v$ (multiple mentions of $v$ in one response count once; 69 cross-area recommendations are excluded from the primary outcome by construction and documented). This yields 1,407,600 venue-level exposure opportunities across 9,214 successes.

\subsection{Predictors, frame, and the unmatched taxonomy}
\label{sec:predictors}

Venue-level predictors operationalize the pre-registered hypotheses from independent sources: Google census fields (rating; log review count; price listed; hours listed; website present), review recency (share of the ten most recent reviews within 90 days, via a commercial scraping API), web-mention volume and review/blog-domain count (web search API), and the Foursquare ladder (open-dataset presence; and among present venues, data-quality rating, tip counts, popularity, via the Places API). Continuous predictors are z-scored; missingness carries explicit indicator terms. Three pre-registered hypotheses (review-text/intent similarity; cross-platform consistency; social presence) were not operationalized --- a data-collection limitation disclosed rather than improvised around. Estimation uses a case-control frame --- all 749 venues carrying a matched wave mention plus 620 controls stratified on area, rating, and review-count bins (1,369 venues, of which 1,275 enter estimation after operational-status and rating-presence restrictions) --- with slope coefficients consistent under outcome-based sampling without weighting \citep{prentice1979logistic}; one covariate (hours listed) was found to encode collection provenance and is excluded from substantive interpretation (Sections~\ref{sec:entry} and~\ref{sec:lessons}).

Unmatched mentions were classified by a web-verified taxonomy (effort allocated by mention count; conservative fabrication bar) into name variants, real venues absent from the frame, out-of-area venues, closed venues, non-venue extractions, and unverifiable names --- the basis of Section~\ref{sec:failure} and of the frame-completion and variant-recovery passes above.

\subsection{Models}
\label{sec:models}

All models are pre-registered in structure; implementation deviations are itemized in the released analysis report. \textbf{M1 (primary):} binomial GLM, successes/trials $\sim$ hypothesis set $+$ engine $+$ persona fixed effects, cluster-robust by venue; Benjamini--Hochberg correction across the nine hypothesis terms. \textbf{M2 (robustness):} GEE with exchangeable within-venue correlation, identical unit and predictors (the pre-registered Bayesian mixed model failed to converge at every tractable scale; the failure and replacement are documented --- 9/9 sign and significance agreement with M1). \textbf{M3:} template-level cluster bootstrap (500 draws) implementing the second pre-registered random effect. \textbf{M4:} the Foursquare ladder among FSQ-present venues. \textbf{M5:} three sensitivity refits --- any-mention outcome; match-score $\geq$95 only; and a refit dropping the Claude arm, which has the lowest mention-matching rate (84.2\%), to check that no headline coefficient depends on the noisiest engine's data (all four engines remain in every primary analysis). \textbf{M7:} conditional logit over within-run choice sets (1,855 runs recommending $\geq$2 matched venues) for the rank-1 margin. A gradient-boosting importance analysis (pre-registered as secondary) was skipped for an environment reason documented in the release. Every analysis is seeded and byte-reproducible: the released scripts regenerate every table and figure identically across runs.

\subsection{Test--retest protocol}
\label{sec:retestproto}

Sixteen queries (one template per persona $\times$ two areas) were re-run on all four engines two weeks post-wave (144 runs) under the identical pipeline; Appendix~\ref{app:retest} compares cross-period venue-set similarity against the within-wave rerun baseline and reports share-of-voice rank stability.

% =====================================================================
\section{Results}
\label{sec:results}

All statistics derive from the frozen analysis snapshot (2026-08-03); models and tables are byte-reproducible from the released analysis scripts. Odds ratios for continuous predictors are per standard deviation of the (log-transformed where noted) predictor; significance statements for the hypothesis set reflect Benjamini--Hochberg correction at FDR 0.05 unless stated otherwise.

\begin{quote}\small
\textbf{Summary of findings.} (1) At least 85.6\% of the 4,776 census venues were never recommended by any of the four systems in 2,208 runs --- 72.6\% even among established venues ($\geq$50 ratings) --- yet visibility among the surfaced is long-tailed, not winner-take-all (leader's share 1.9\%; Gini 0.668). (2) Entry into answers is associated with documentation --- review volume (\OR{1.64}), an own website (\OR{1.92}), listed price (\OR{1.54}), web mentions (\OR{1.44}) --- while star rating is null at entry (\OR{0.89}) and significant only for ranking first within an answer (\OR{1.17}). (3) Foursquare presence and standing predict nothing positive at either margin. (4) The dominant failure mode is staleness (93 closed-venue recommendations), not fabrication (0.08\% of valid mentions). (5) Answers churn run-to-run (Jaccard 0.22--0.45) and across engines (top-20 Jaccard 0.33--0.54), but a two-week test--retest shows no drift beyond that noise floor.
\end{quote}

\subsection{Collected corpus and extraction quality}
\label{sec:corpus}

The confirmatory wave comprises 2,208 search-grounded runs (Perplexity 960, Gemini 480, OpenAI 480, Claude 288) across 96 persona-conditioned queries and seven collection days, yielding 12,439 valid venue mentions after rule-based filtering. Mention extraction was validated by independent double annotation with human adjudication (mention-level precision 97.6\%, recall 99.1\%; strict run-level agreement 91.5\% after remediation), and entity matching by a stratified 100-mention audit ($\approx$98--99\% population-weighted correct-entity rate). After matcher repair and evidence-audited recovery of name variants, 87.9\% of valid mentions resolve to a registry venue; per-engine match rates range from 84.2\% (Claude) to 91.4\% (Gemini). The remainder decomposes in Section~\ref{sec:failure}.

\subsection{Most of the market is invisible}
\label{sec:invisible}

\begin{figure}[t]
\centering
\includegraphics[width=\linewidth]{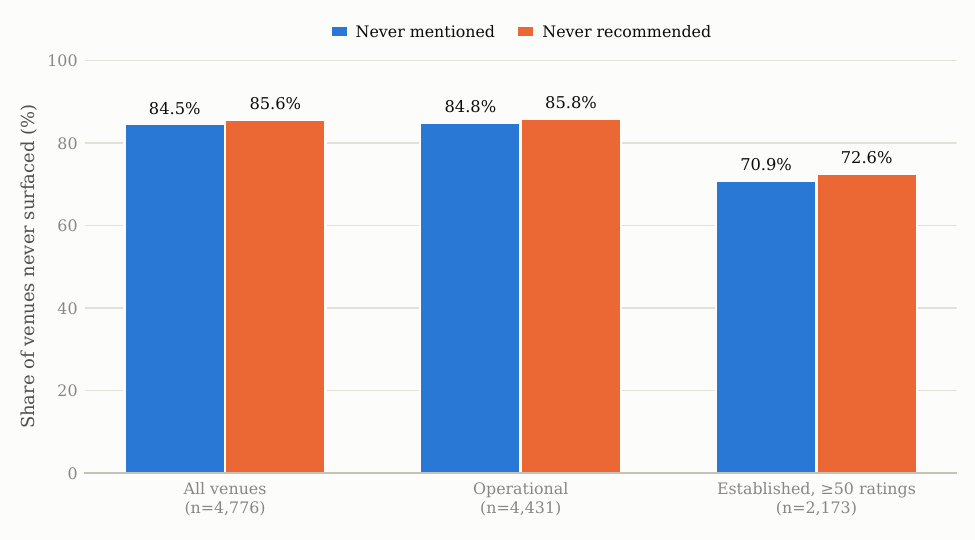}
\caption{Invisibility by denominator. Share of census venues never mentioned and never recommended by any of the four systems across all 2,208 runs, under three nested frames: all venues, confirmed-operational venues, and established venues with $\geq$50 Google ratings. All rates are floors (Appendix~\ref{app:census}).}
\label{fig:invisibility}
\end{figure}

\begin{figure}[t]
\centering
\includegraphics[width=\linewidth]{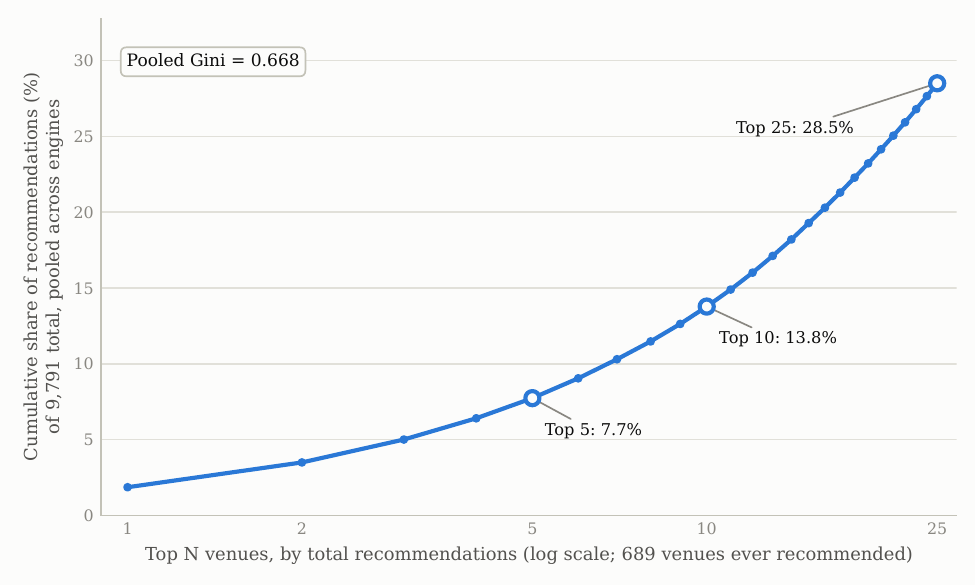}
\caption{Concentration among the visible. Cumulative share of all 9,791 pooled recommendations captured by the top-$N$ venues. The leading venue holds 1.9\%; the top 25 venues jointly hold 28.5\%. Severe entry filtering (Figure~\ref{fig:invisibility}) coexists with a long tail rather than winner-take-all concentration.}
\label{fig:concentration}
\end{figure}

The 2,208 runs --- each a single persona query answered by one system with a short list of venues --- produced 9,791 venue recommendations in total, and those recommendations collectively cover only 689 of the 4,776 census venues. We term the share of venues never recommended by any system the \textbf{invisibility rate}: here \textbf{85.6\%} (4,087 of 4,776 venues; Figure~\ref{fig:invisibility}) --- a floor rather than an estimate, since any venue absent from the census frame is, by the adversarial-probe analysis of Appendix~\ref{app:census}, almost surely never recommended; under the broadest defensible frame the rate exceeds 92\%. The rate is nearly unchanged among confirmed-operational venues (85.8\%) and remains 72.6\% among established venues with at least fifty Google ratings --- the denominator least chargeable with including marginal businesses. Visibility, however, is not winner-take-all: the most-recommended venue (Seniman Coffee, Ubud) accounts for only 1.9\% of all recommendations, the pooled Gini coefficient over recommended venues is 0.67, and the top five venues jointly capture 7.7\% of recommendations (top 25: 28.5\%) (Figure~\ref{fig:concentration}). The picture is a severe entry filter followed by a comparatively open field among the visible: being recommended at all is the scarce event; no small clique monopolizes the answers.

\subsection{The entry margin: documentation, not rating}
\label{sec:entry}

\begin{figure}[t]
\centering
\includegraphics[width=\linewidth]{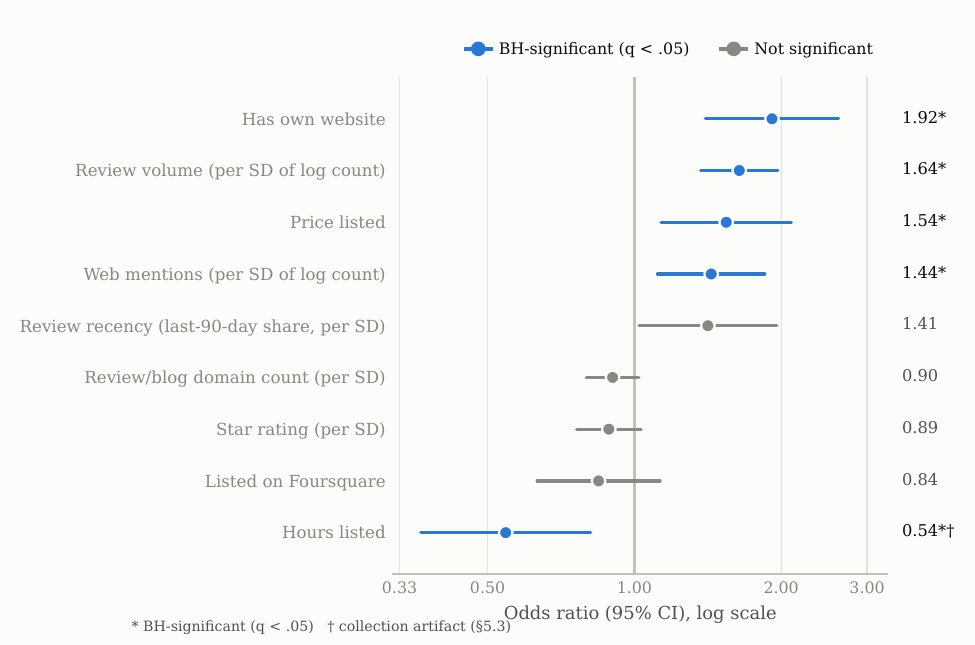}
\caption{Entry-margin factor estimates (M1). Odds ratios with 95\% confidence intervals, log scale; filled blue markers are Benjamini--Hochberg-significant at FDR 0.05. The hours-listed coefficient is a collection-provenance artifact (Section~\ref{sec:entry}).}
\label{fig:forest}
\end{figure}

Table~\ref{tab:entry} (Figure~\ref{fig:forest}) reports the pre-registered binomial GLM for the probability that a venue is recommended in an eligible run (venue $\times$ persona $\times$ engine; cluster-robust by venue; engine and persona fixed effects). Four factors survive correction, and they share a theme:

\begin{itemize}
\item \textbf{Review volume}: \OR{1.64} per SD of log review count ($p < .001$).
\item \textbf{An own website}: \OR{1.92} ($p < .001$) --- the largest entry-margin effect.
\item \textbf{Listed price information}: \OR{1.54} ($p = .010$ adj.).
\item \textbf{Web-mention volume}: \OR{1.44} ($p = .010$ adj.).
\end{itemize}

Star rating is null at this margin (\OR{0.89}, adj.\ $p = .14$): conditional on volume and documentation, a venue's average rating shows no detectable association with whether AI systems surface it at all. Review recency leans positive (\OR{1.41}) but narrowly misses the corrected threshold (adj.\ $p = .054$); we report it as suggestive. Foursquare presence --- the data-provider hypothesis this study was partly designed to test --- is null (\OR{0.84}, adj.\ $p = .25$), and the graded ladder within Foursquare-listed venues (data-quality rating, tips, popularity; Table~\ref{tab:ladder}) is uniformly non-significant. One covariate carries a collection artifact we identified post hoc and exclude from substantive interpretation: venues with listed opening hours appear \emph{less} likely to be recommended (\OR{0.54}, adj.\ $p = .009$), but all 70 venues added to the census via mention-driven lookup --- venues that are recommended nearly by construction --- lack the hours field because the lookup request omitted it. The hours covariate thus partially encodes registry provenance rather than profile completeness; we report it for transparency and discuss the general lesson --- audit covariates can inherit structure from how the auditor discovered the entities --- in Sections~\ref{sec:lessons} and~\ref{sec:limitations}. Robustness: a GEE with exchangeable within-venue correlation agrees with the primary model on sign and significance for all nine hypothesis terms; template-level cluster bootstrap (500 draws) leaves every point estimate essentially unmoved; and all conclusions are stable under three sensitivity refits (any-mention outcome; matches $\geq$95 only; excluding the Claude arm, whose search budget was capped).

\begin{table}[t]
\centering
\small
\caption{Entry margin (M1): pre-registered binomial GLM for the probability a venue is recommended in an eligible run. Unit: venue $\times$ persona $\times$ engine (1,407,600 exposure opportunities; 9,214 successes; 1,275 venues). Engine and persona fixed effects included; standard errors cluster-robust by venue; $p$-values Benjamini--Hochberg-corrected across the nine hypothesis terms. Continuous predictors per SD.}
\label{tab:entry}
\begin{tabular}{lccc}
\toprule
Factor & OR & 95\% CI & adj.\ $p$ \\
\midrule
Has own website & 1.92 & [1.40, 2.62] & $<.001$ \\
Review volume (log count, per SD) & 1.64 & [1.37, 1.97] & $<.001$ \\
Price listed & 1.54 & [1.13, 2.10] & .010 \\
Web mentions (log count, per SD) & 1.44 & [1.12, 1.85] & .010 \\
Review recency (last-90-day share, per SD) & 1.41 & [1.02, 1.96] & .054 \\
Review/blog domain count (per SD) & 0.90 & [0.80, 1.02] & .134 \\
Star rating (per SD) & 0.89 & [0.76, 1.03] & .135 \\
Listed on Foursquare & 0.84 & [0.63, 1.13] & .252 \\
Hours listed$^\dagger$ & 0.54 & [0.36, 0.81] & .009 \\
\bottomrule
\multicolumn{4}{l}{\footnotesize $^\dagger$Collection-provenance artifact; excluded from interpretation (see text).}
\end{tabular}
\end{table}

\begin{table}[t]
\centering
\small
\caption{The Foursquare ladder (M4): graded standing within the 480 Foursquare-listed venues in the case-control set, added to the M1 specification. Uncorrected $p$-values; all three ladder terms are non-significant.}
\label{tab:ladder}
\begin{tabular}{lccc}
\toprule
Foursquare signal & OR & 95\% CI & $p$ \\
\midrule
Data-quality (veracity) rating (per SD) & 1.12 & [0.88, 1.43] & .352 \\
Tip count (log, per SD) & 1.17 & [0.95, 1.45] & .133 \\
Popularity score (per SD) & 1.12 & [0.90, 1.40] & .308 \\
\bottomrule
\end{tabular}
\end{table}

\subsection{The ranking margin: rating returns}
\label{sec:rank}

\begin{figure}[t]
\centering
\includegraphics[width=\linewidth]{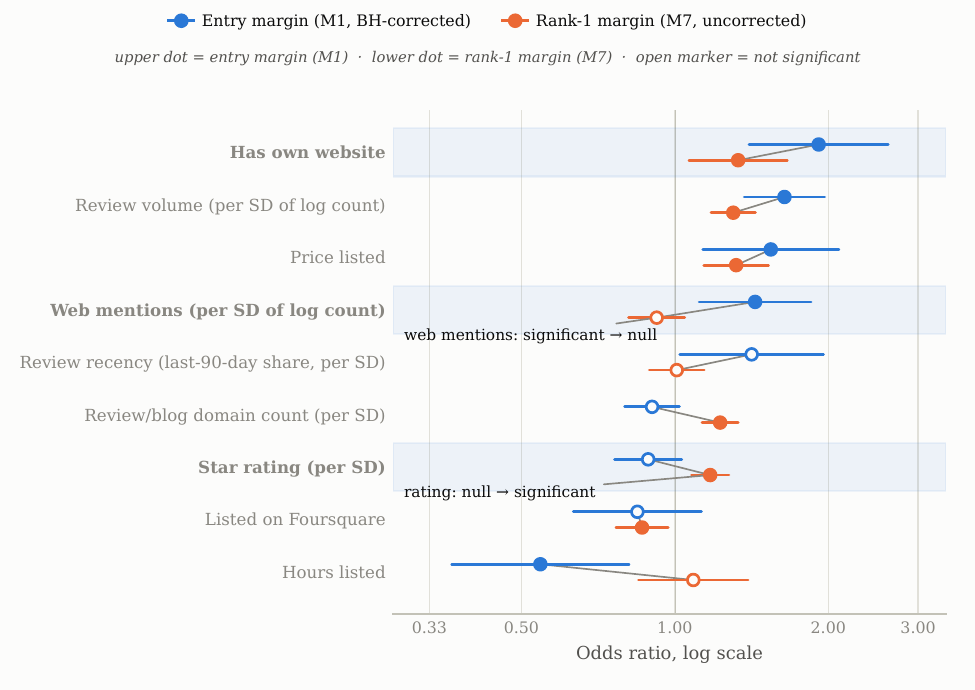}
\caption{The two-margin structure. Entry-margin (M1, BH-corrected) and rank-1-margin (M7, uncorrected) odds ratios for the same factors. Highlighted rows mark the two dissociations: web mentions matter for entry but not rank; rating matters for rank but not entry.}
\label{fig:margins}
\end{figure}

The entry results invert once we ask which venue is named \emph{first}. Among the 1,855 runs recommending at least two matched venues, a conditional logit over within-run choice sets (Table~\ref{tab:rank}, Figure~\ref{fig:margins}) shows rating significantly predicting first position (\OR{1.17} per SD, $p = .0002$), alongside review volume (\OR{1.30}). The documentation variables weaken sharply at this margin: web-mention volume is null within choice sets (\OR{0.92}), and the website effect attenuates from 1.92 at entry to a nominal 1.33 (M7 carries no pre-registered correction family; within-set significance is labeled by 95\% CI only). The count of review/blog domains covering a venue --- null at entry --- is positively associated with first position (\OR{1.23}), consistent with editorial coverage mattering for ordering rather than admission. Foursquare presence is, if anything, nominally \emph{negative} within choice sets (\OR{0.86}, 95\% CI 0.77--0.97) --- still no support for the hypothesized positive effect; and because this margin conditions on entry, within-set coefficients of entry-relevant variables can carry selection artifacts, so we do not interpret the negative sign. \textbf{AI venue visibility is two-margined: documentation and volume dominate whether a venue enters the answer; rating --- null at entry --- emerges only in ordering the venues that made it in.} This reconciles our entry-margin null for rating with conjoint evidence that rating dominates within fixed choice sets \citep{baig2026hotel}: the two designs measure different margins, and both are right about their own.

\begin{table}[t]
\centering
\small
\caption{Rank-1 margin (M7): conditional logit over within-run choice sets (1,855 runs recommending $\geq$2 matched venues; 8,450 venue-run alternatives). Outcome: being the run's first-named recommendation. Uncorrected $p$-values; M7 carries no pre-registered correction family.}
\label{tab:rank}
\begin{tabular}{lccc}
\toprule
Factor & OR & 95\% CI & $p$ \\
\midrule
Star rating (per SD) & 1.17 & [1.08, 1.28] & $<.001$ \\
Review volume (log count, per SD) & 1.30 & [1.18, 1.44] & $<.001$ \\
Review recency (last-90-day share, per SD) & 1.01 & [0.89, 1.14] & .890 \\
Has own website & 1.33 & [1.07, 1.66] & .012 \\
Price listed & 1.32 & [1.14, 1.53] & $<.001$ \\
Hours listed & 1.09 & [0.85, 1.39] & .509 \\
Listed on Foursquare & 0.86 & [0.77, 0.97] & .014 \\
Web mentions (log count, per SD) & 0.92 & [0.81, 1.05] & .206 \\
Review/blog domain count (per SD) & 1.23 & [1.13, 1.33] & $<.001$ \\
\bottomrule
\end{tabular}
\end{table}

\subsection{Stability}
\label{sec:stability}

\begin{figure}[t]
\centering
\includegraphics[width=0.72\linewidth]{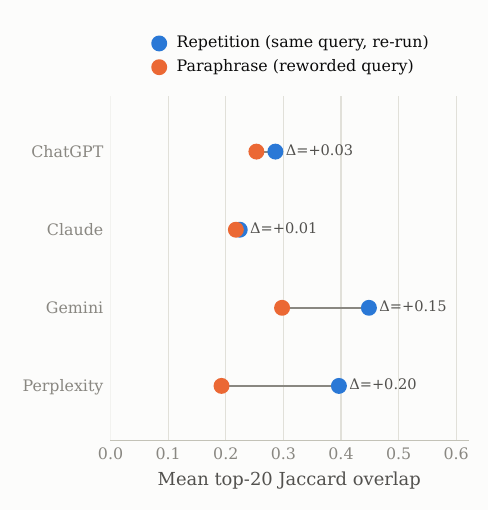}
\caption{Repetition vs.\ paraphrase stability. Mean top-20 Jaccard overlap between venue sets for identical repeated queries (repetition) and meaning-preserving rewordings (paraphrase), per engine. Paraphrase similarity is lower for every engine.}
\label{fig:stability}
\end{figure}

Identical queries repeated on the same engine return substantially different venue sets (mean Jaccard: Gemini 0.45, Perplexity 0.40, OpenAI 0.29, Claude 0.22). Meaning-preserving paraphrases perturb results further: paraphrase similarity falls below repetition similarity for all four engines, pronouncedly for Perplexity (0.19 vs 0.40) and Gemini (0.30 vs 0.45) and minimally for Claude ($\Delta = 0.01$ at its smaller repetition count), replicating the paraphrase-brittleness pattern reported for commercial RAG systems \citep{jack2026paraphrase} (Figure~\ref{fig:stability}). A pre-registered test--retest holdout (16 queries $\times$ 4 engines re-run two weeks post-wave, $n = 144$) finds cross-period venue-set similarity comparable to the same-period rerun baseline for every engine (pooled Jaccard 0.375; share-of-voice Spearman $r = 0.47$ under unequal-arm attenuation; Appendix~\ref{app:retest}): answer churn reflects sampling stochasticity, not temporal drift, supporting the treatment of wave-period visibility as a persistent venue property.

\subsection{Systems disagree on who to recommend}
\label{sec:disagree}

\begin{figure}[t]
\centering
\includegraphics[width=0.62\linewidth]{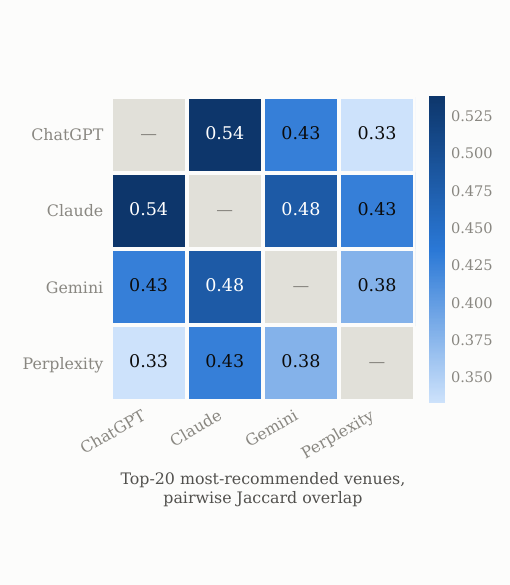}
\caption{Cross-engine agreement. Pairwise Jaccard overlap between the four engines' top-20 most-recommended venue sets.}
\label{fig:jaccard}
\end{figure}

\begin{figure}[t]
\centering
\includegraphics[width=\linewidth]{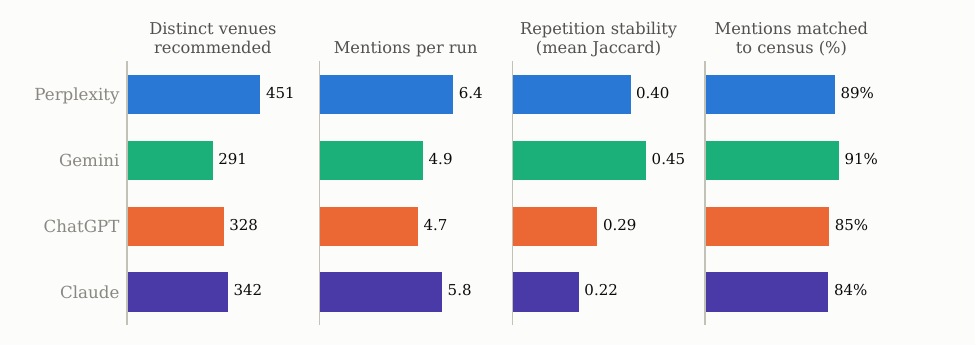}
\caption{Engine fingerprints. Distinct venues recommended, mention volume per run, repetition stability (mean Jaccard overlap between venue sets of identical repeated queries), and share of mentions matched to the census, per system. Perplexity recommends the broadest slate; Gemini is the most stable and best-matched; Claude is the least stable (its search budget was capped at two searches per run, Section~\ref{sec:systems}).}
\label{fig:fingerprint}
\end{figure}

\begin{figure}[t]
\centering
\includegraphics[width=\linewidth]{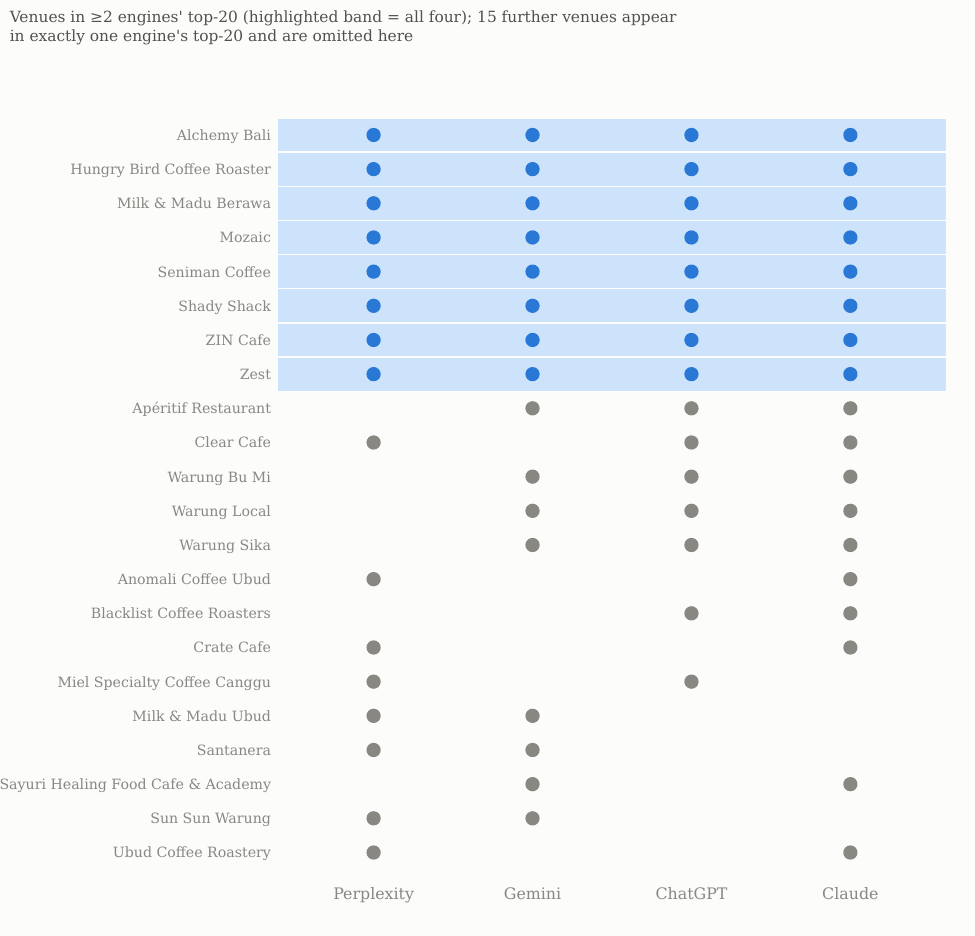}
\caption{The consensus core. Top-20 membership for every venue appearing in at least two engines' top-20 lists; the highlighted band marks the eight venues in all four. Fifteen further venues appear in exactly one engine's top-20 and are omitted.}
\label{fig:consensus}
\end{figure}

Pairwise agreement between engines' top-20 recommended sets ranges from Jaccard 0.33 (Perplexity--OpenAI) to 0.54 (OpenAI--Claude) (Figure~\ref{fig:jaccard}). Only 8 venues appear in all four engines' top-20; 15 appear in exactly one (Figure~\ref{fig:consensus} maps the full membership pattern). This extends the low cross-model agreement documented for brand recommendation (41.6\% top-brand agreement; \citealp{zatuchin2026owns}) to the venue domain: optimizing visibility for one assistant is not optimizing for ``AI'' in general. The disagreement sits within broader behavioral differences between the systems --- breadth of venues recommended, answer verbosity, rerun stability, and census match rate --- summarized side by side in Figure~\ref{fig:fingerprint}.

\subsection{What the systems read}
\label{sec:sources}

\begin{figure}[t]
\centering
\includegraphics[width=\linewidth]{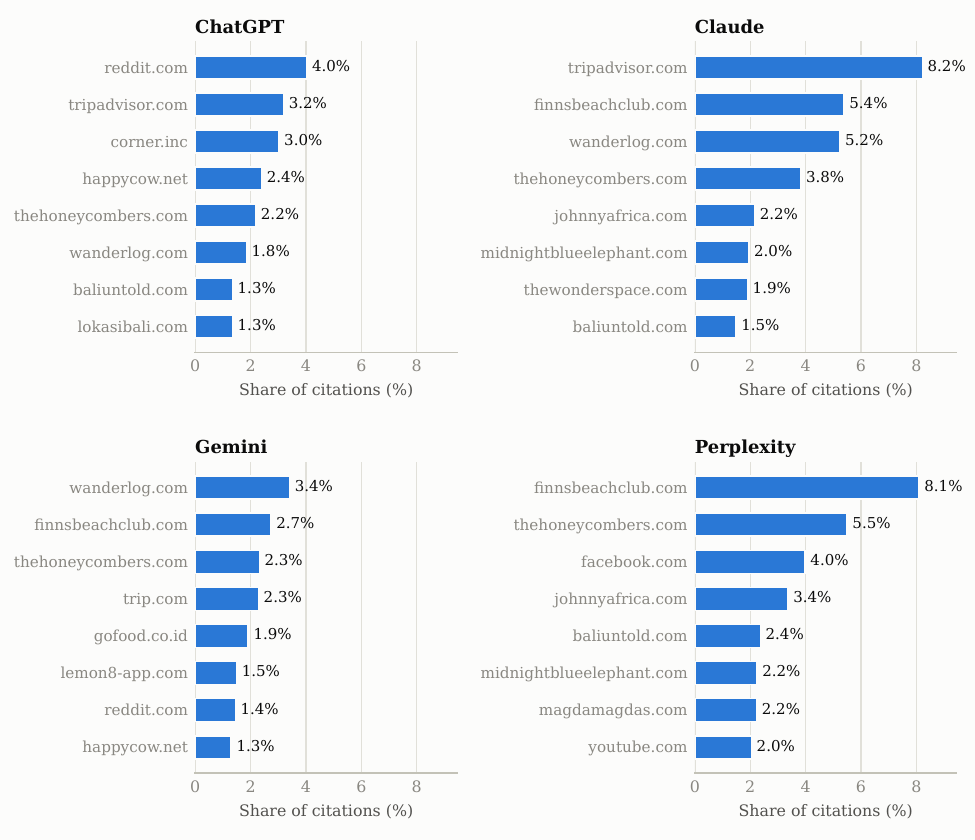}
\caption{Citation diets. Top cited domains by share of each engine's grounding citations. Gemini's API exposes only grounding-redirect URLs; its domains were recovered from citation titles by deterministic mapping, and unrecoverable citations are excluded from its panel.}
\label{fig:domains}
\end{figure}

The 26,993 grounding citations attached to wave runs span 986 domains (Figure~\ref{fig:domains}). Pooled, the most-cited domain is not a platform but a single venue's own website (finnsbeachclub.com, 5.3\%) --- an in-the-wild demonstration that venue-owned content can out-cite TripAdvisor (2.3\%) --- followed by regional listicle and travel-blog domains (thehoneycombers.com 3.9\%, wanderlog.com 2.2\%), with community sources (Reddit 1.3\%, YouTube 1.3\%) and review platforms making up the remainder. Engines differ sharply in source diets (Figure~\ref{fig:domains}, per-engine panels); we note a measurement caveat for Gemini, whose API reports only a grounding-redirect domain and whose true sources were recovered from citation titles (see the caption of Figure~\ref{fig:domains}).

\subsection{Failure modes: staleness, not fabrication}
\label{sec:failure}

\begin{figure}[t]
\centering
\includegraphics[width=\linewidth]{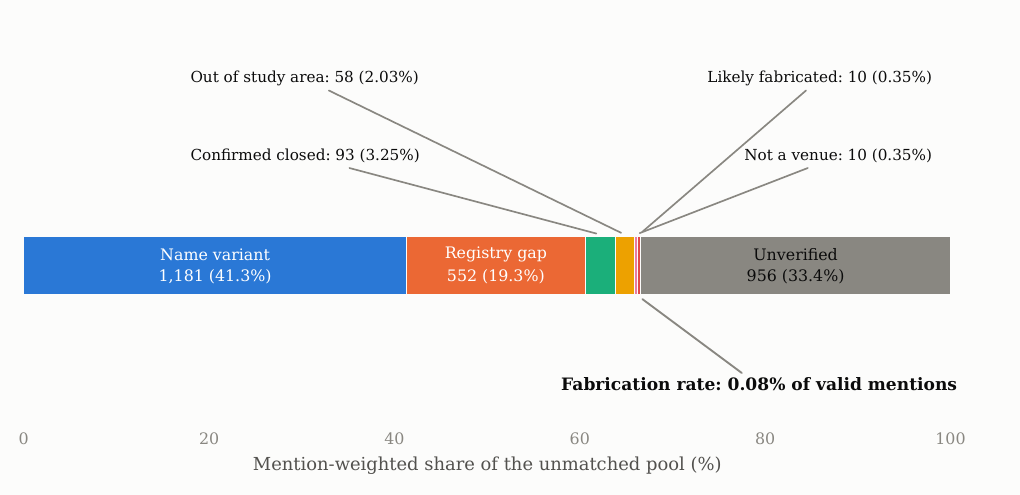}
\caption{Decomposition of the confirmatory wave's pre-recovery unmatched pool (2,452 mentions), by web-verified taxonomy class. Recovered name variants and an unresolved long tail dominate; likely fabrication is 10 mentions (0.08\% of all valid wave mentions), while confirmed-closed venues account for 93 recommendations.}
\label{fig:taxonomy}
\end{figure}

Before variant recovery, 19.7\% of valid wave mentions (2,452) lacked a census match. Every unmatched name was classified by a per-name, web-verified audit conducted over the study's full pre-recovery corpus; applying its labels to the wave pool (Figure~\ref{fig:taxonomy}): 38.4\% (941 mentions) are name variants of registered venues --- rebrands, nicknames, transliterations --- recovered through 142 evidence-audited mappings, 3.8\% are venues verified permanently closed, 2.4\% are real venues outside the study polygons, and the 54.6\% remainder is unverifiable long-tail singletons plus variant candidates below the recovery evidence bar, left unmatched; 1,511 mentions (12.1\% of valid mentions) remain unresolved after recovery. Venues the systems surfaced that food-type enumeration had missed were resolved into the census during frame construction (Section~\ref{sec:censusframe}) and therefore do not appear in this pool. Genuine fabrication is rare to the point of near-absence: a single name (10 mentions, \textbf{0.08\% of all valid mentions}) survived conservative scrutiny as likely invented. The practically significant failure mode is staleness: systems recommended permanently closed venues in \textbf{93 mentions across 14 confirmed-closed establishments}, and our census-based design detects these only because closure was independently verifiable --- a sampled audit would count most of them as successes. For local discovery, the risk is not that AI invents restaurants; it is that AI remembers restaurants that no longer exist.

% =====================================================================
\section{Discussion}
\label{sec:discussion}

\subsection{A two-margin account of AI venue visibility}
\label{sec:twomargin}

Our central empirical pattern is a dissociation. Whether a venue enters an AI-generated answer is associated with the paper trail it has accumulated --- review volume, an own website, listed price data, third-party mentions --- while its star rating carries no detectable weight. Yet among venues that do enter, rating is the strongest ordering signal we measure. We read this through the retrieval-then-generation architecture of the systems audited: entry is gated upstream, where a venue must first exist in the sources the model retrieves --- listicles, review aggregators, its own site --- and the volume-and-documentation variables are precisely the ones that determine whether those sources have anything to say. Ordering happens downstream, where the model composes an answer from retrieved candidates and can attend to quality cues like rating. The dissociation also dissolves an apparent conflict in the literature: conjoint audits, which place every candidate before the model by construction \citep{baig2026hotel}, find rating dominant --- and so do we, at that margin. Audits with a real market denominator find long-tail invisibility that no within-set experiment can observe \citep{jack2026prominence}, and our census sharpens this to a population rate. Both traditions are measuring real, different stages of one pipeline.

The practical inversion is stark: the factor local businesses invest most heavily in signaling --- the rating --- does nothing for discovery, and the factors that govern discovery are mundane infrastructure. A 4.9-star caf\'e with thin documentation is, to these systems, indistinguishable from absent.

\subsection{What the null results mean}
\label{sec:nulls}

The Foursquare null deserves plain statement because we designed part of this study to find the opposite. Presence in the open POI dataset --- and, among present venues, its quality tier, engagement counts, and popularity signal --- shows no positive association with visibility at either margin once documentation and volume are controlled (the within-answer rank coefficient is, if anything, nominally negative --- an entry-conditioned estimate we do not interpret, Section~\ref{sec:rank}). Two readings are compatible with our data: assistants' retrieval simply does not touch this dataset for venue discovery in our market, or its signal is redundant with the web presence we already measure. Either way, the folk practice of treating POI-dataset inclusion as an ``AI visibility hack'' finds no support in the first direct test. We note the history of this coefficient within our own study (Section~\ref{sec:lessons}) as a caution for anyone re-estimating it with looser matching.

Rating's entry-margin null merits equal care in the other direction: it does not say quality is irrelevant --- rating operates at the ordering stage, and review \emph{volume} (which correlates with quality mechanisms we do not isolate) dominates entry. The claim licensed by our design is narrower and more useful: improving a rating from 4.3 to 4.7 without growing the documentation trail should not be expected to change whether AI systems surface the venue at all.

\subsection{Instability without drift}
\label{sec:instability}

Repeated identical queries return substantially different venue sets, and paraphrases perturb them further --- yet the two-week test--retest shows cross-period similarity no worse than the same-day rerun baseline. The churn is stochastic, not temporal: these systems behave less like a ranked index that updates and more like a noisy sampler over a persistent visibility distribution. Combined with low cross-engine agreement (top-20 Jaccard 0.33--0.54), this has a concrete implication for the emerging GEO industry: single-shot, single-engine ``visibility checks'' --- the standard product offering --- measure noise. Meaningful measurement requires repetition across engines and phrasings, which is an argument for audit methodology \citep{jack2026paraphrase} arriving intact in commercial practice.

\subsection{Staleness, not hallucination}
\label{sec:stalediscussion}

The systems almost never invented venues: one unverifiable name in more than 12,000 valid mentions. They recommended permanently closed venues 93 times. This inverts the popular framing of AI risk in local discovery. The generation layer, grounded in retrieval, is faithful to its sources; the sources are stale. A closed caf\'e's reviews, listicle entries, and blog mentions persist --- precisely the signals our entry-margin model rewards --- so the documentation trail that creates visibility also sustains it after death. For platforms, the implication is that closure signals propagate far more slowly than reputation signals; for consumers, that the failure mode to expect is not fiction but anachronism.

\subsection{Methodological lessons: the audit is part of the instrument}
\label{sec:lessons}

Two of our findings are about auditing itself. First, factor-importance conclusions are sensitive to entity-matching quality: our data-provider coefficient moved from a strong positive (\OR{1.70}) to a null under a matching remediation motivated entirely by validation evidence, because matching errors correlated with the predictor. No prior audit in this literature validates its matching pipeline; ours suggests they must. Second, covariates can inherit structure from the auditor's own collection path: our hours-listed variable partially encoded \emph{how a venue entered our census} (mention-driven lookup omitted the hours field), manufacturing a spurious negative association that survived multiple-comparison correction. Both errors were caught by pre-registered validation gates rather than luck; we report them as costs of honest measurement and as evidence that observational AI audits need the same instrument-validation discipline as survey research.

\subsection{Implications}
\label{sec:implications}

For independent venues, the actionable hierarchy our estimates support is: be documented before being excellent --- maintain an own website, keep platform profiles complete, accumulate review volume and third-party mentions; the rating begins to matter once you are being retrieved. For platform and assistant builders: the 85.6\% invisible majority is not a quality filter --- it includes three-quarters of \emph{established} venues --- and staleness propagation is the clearest correctable harm. For researchers, the census-denominated design is portable to any bounded market and is, we argue, the only way to measure what sampled audits structurally cannot: who never appears at all.

Review-text content --- whether detailed, attribute-rich reviews causally improve retrieval --- is the natural next hypothesis; it requires full review corpora and ideally an intervention design, and we leave it to a dedicated study.

% =====================================================================
\section{Limitations}
\label{sec:limitations}

\paragraph{Observational design.} Our factor estimates are associations under controls, not causal effects. The website coefficient illustrates the general caution: established, professionally run venues both maintain websites and accumulate digital footprint, so unmeasured scale or professionalism could contribute to the association even though it survives controls for review volume and web mentions --- our best available scale proxies. The within-choice-set rank margin (M7) is likewise conditional on entry. Causal versions of these questions require attribute manipulation (as in conjoint designs) or venue-side interventions over time; we regard the latter as the natural follow-up.

\paragraph{Frame dependency and market churn.} The census enumerates the Google-listed food-service market (Section~\ref{sec:censusframe}, Appendix~\ref{app:census}); venues outside that frame --- unlisted micro-warungs, atypically typed outlets --- are underrepresented, and Bali's venue turnover means the census is a snapshot. Both gaps bias the invisibility rate downward, so headline rates are floors; factor estimates apply to the listed market, which is also the market the audited systems retrieve from.

\paragraph{Access route and configuration.} We audit search-grounded APIs, not the consumer apps; provider-side serving configurations are opaque and can change without notice \citep{bouchaud2024auditing}. One configuration is ours: Claude's search budget was capped at two per run; the cap is disclosed, it was binding in practice (97.6\% of Claude wave runs consumed both searches), and all conclusions survive removing the Claude arm (Section~\ref{sec:models}, M5).

\paragraph{Single region, single language.} Two adjacent Balinese submarkets, English queries, a tourist/nomad demand profile. Linguistic and geographic generalization is untested here and is documented to matter elsewhere \citep{venkateswaran2026linguistic, bhagat2024richer}; replication in a structurally different market (our next study region) is required before generalizing.

\paragraph{Measured instrument error.} Extraction and matching carry known, quantified error (strict run-level 91.5\%; matching $\approx$98--99\% population-weighted) with error classes reported; one covariate (hours listed) encoded collection provenance and is excluded from interpretation --- we flag covariate--provenance contamination as a general audit hazard. Three pre-registered predictors (review-text/intent similarity, cross-platform consistency, social presence) were not operationalized.

\paragraph{Unequal arms and temporal scope.} Repetition counts differ by engine (budget-driven, pre-registered); per-engine precision differs accordingly. The wave spans seven days plus a two-week retest --- stability beyond that horizon, and across model generations, is unmeasured.

% =====================================================================
\section{Ethics and Disclosure}
\label{sec:ethics}

\paragraph{Funding and competing interests.} This study was funded and conducted by Norly (norly.co), a company selling review-management and AI-visibility tools to hospitality businesses --- a commercial interest directly adjacent to the research question. We disclose this prominently and describe the insulation measures rather than asking for trust: the hypotheses, design, and analysis plan were pre-registered publicly before confirmatory collection; the analysis dataset was frozen before models ran, with every subsequent change documented; validation gates were adjudicated against written rules with failures reported (Section~\ref{sec:method}); and the study's results include prominent null findings against commercially convenient hypotheses --- notably the Foursquare data-provider test the study was partly designed around --- as well as a rating-null result that contradicts common industry messaging. Marketing derivative works are bound to quote only effect sizes published here.

\paragraph{Human subjects and privacy.} The study involves no human subjects and no personal data. Review text was processed only in aggregate to compute venue-level recency statistics; no review text, reviewer identity, or other user-generated content is released. Venue-level facts (names, locations, ratings, counts) are business information already public on the platforms concerned.

\paragraph{Data collection posture.} AI systems were queried through public, paid APIs under their terms of service. Venue-side features derive from platform APIs and licensed third-party collection services; scraped material was used solely as analysis input (predictor variables), never redistributed. The released replication package contains derived per-venue features, aggregate tables, the full query instrument, collection and analysis code, and the pre-registration --- not raw platform content.

\paragraph{Potential harms considered.} Named venues appear only in aggregate visibility statistics computed from public AI outputs; we report no venue-level quality judgments of our own. The taxonomy identifies closed venues from public signals; misclassification risk was controlled by a conservative evidence bar with per-item sources retained. We considered whether publishing factor estimates could enable manipulation of AI recommendation systems; the factors identified (documentation, review volume, web presence) are the same legitimate signals platforms already encourage businesses to maintain, and the manipulation-specific literature is already public \citep{kumar2024manipulating}.

% =====================================================================
\clearpage
\bibliography{references}

% =====================================================================
\appendix

\section{Test--retest stability}
\label{app:retest}

Per the pre-registered protocol (Section~\ref{sec:retestproto}), 16 queries (one template per persona $\times$ two areas) were re-run on all four engines two weeks after the confirmatory wave: 144 runs (Perplexity 3 repetitions, others 2), zero collection errors, processed by the identical frozen pipeline (no registry changes; deterministic re-matching reproduces wave assignments).

\begin{table}[h]
\centering
\small
\caption{Cross-period venue-set similarity (wave vs.\ retest, same query $\times$ engine) against the within-wave repetition baseline.}
\label{tab:retest}
\begin{tabular}{lcc}
\toprule
Engine & Cross-period Jaccard & Within-wave rerun Jaccard \\
\midrule
Gemini & 0.593 & 0.449 \\
Perplexity & 0.344 & 0.396 \\
OpenAI & 0.309 & 0.286 \\
Claude & 0.254 & 0.224 \\
\midrule
Pooled & 0.375 & --- \\
\bottomrule
\end{tabular}
\end{table}

Cross-period similarity is statistically comparable to the same-period rerun baseline for every engine (Gemini's cross-period figure exceeds its baseline because the cross-period comparison aggregates over more wave runs). Two weeks of elapsed time adds no detectable instability beyond run-to-run sampling noise.

Share-of-voice stability: among the 359 venues surfaced on these queries in either period, the Spearman correlation between wave and retest recommendation rates is $r = 0.47$, and 47\% of venues surfaced in both periods. This correlation is attenuated by design: the retest arm (144 runs) is $\sim$2.6$\times$ smaller than the wave arm on the same queries (368 runs), so sampling error alone bounds the attainable correlation well below 1. The Jaccard-versus-baseline comparison above is therefore the cleaner stability statement. Together they support the paper's treatment of wave-period visibility as a persistent venue property measured through a stochastic channel (Section~\ref{sec:instability}).

\section{Census completeness}
\label{app:census}

The census is a complete enumeration of an operationally defined frame --- the Google-listed food-service market within the study polygons (Section~\ref{sec:censusframe}) --- not of a metaphysical market. This appendix quantifies what the frame misses and shows that every measurable gap biases the headline invisibility rate downward.

\paragraph{Dual-frame capture--recapture.} We compared the census against an independent enumeration: Foursquare open-data ``Dining and Drinking'' places within the same polygons. A naive Lincoln--Petersen/Chapman estimate using strict name matching is dominated by artifacts (Foursquare internal duplicates, stale entries, and matcher false-negatives) and is not usable. We therefore corrected all three: deduplicating Foursquare rows ($-$2\%), restricting to entries refreshed since 2024, and correcting matcher recall by hand-reviewing a random 120-row sample of ``unmatched'' Foursquare rows --- 33\% of which are in fact census venues under relaxed matching. The corrected overlap ($\approx$1,789 of 3,516 fresh Foursquare rows) yields a Chapman estimate of $\hat N \approx 9{,}400$ ``any-list'' food entities, implying census coverage of roughly 51\% of that broadest frame. This is conservative in both directions we can check: the relaxed matcher itself has imperfect recall (inflating $\hat N$), and Foursquare freshness does not imply operating status (inflating $\hat N$ again).

\paragraph{What the non-overlap contains.} A qualitative hand audit of the Foursquare-only stratum finds it dominated by entities of a different kind from the study's query market: micro-warungs and stalls with no or atypical Google presence, hotel sub-outlets (rooftop bars, in-house counters), non-venue entities (a cooking school), user-typo artifacts, and edge-of-polygon fringe. These are overwhelmingly absent from the audited systems' retrieval substrate as well.

\paragraph{The adversarial probe.} The audited systems are themselves the strongest available probe of denominator adequacy: any venue an engine can surface that the census lacks would appear as an unresolved mention. Across all 16,447 valid mentions collected by the study's pipeline (pilot, wave, and retest), 87.8\% resolve to the census, and the unmatched-mention taxonomy (Section~\ref{sec:failure}) decomposes the remainder into name variants of census venues, out-of-area venues, closed venues, and unverifiable long-tail singletons. Probing 119 AI-surfaced candidate names against Places Text Search yielded only 57 venues the census had lacked (109 resolved to existing entries); the engines never named the Foursquare-only micro-warung stratum. For measuring AI visibility, the census is demonstrably near-complete \emph{with respect to the universe AI systems draw from} --- which is the universe the numerator lives in.

\paragraph{Direction of residual error.} Any venue missing from the denominator is, by the probe above, almost surely never recommended; adding such venues raises the invisibility rate. Under the corrected broader frame ($\hat N \approx 9{,}400$), the never-recommended share is at least 92.6\%. The reported invisibility rates (85.6\% overall; 72.6\% among established venues) are therefore floors: every completeness critique makes the headline stronger, not weaker.

\section{The query instrument}
\label{app:queries}

The complete instrument: eight personas $\times$ six templates. Each template
carries an \textsf{\{area\}} slot instantiated as ``Canggu, Bali'' and
``Ubud, Bali'', yielding the 96 unique queries of the confirmatory wave,
reproduced verbatim below. The instrument was designed by the research team
at the start of the project, anchored on a real traveler-style query, and
frozen at pre-registration before any confirmatory run; templates within a
persona are deliberate paraphrases of a constant intent (Section~\ref{sec:instrument}).

\small

\subsection*{Digital nomad --- place to work}
\begin{itemize}\itemsep2pt
\item \textit{nomad\_1}: I'm 35, just arrived in \textsf{\{area\}} and I'll be staying for about a month. I work remotely and I'm looking for a cafe where I can work on my laptop for a few hours --- good wifi, decent coffee, and it shouldn't be too loud. Where should I go?
\item \textit{nomad\_2}: Best laptop-friendly cafes in \textsf{\{area\}}? I need stable wifi and power outlets, and I usually sit for half a day.
\item \textit{nomad\_3}: I'm a digital nomad based in \textsf{\{area\}} this month. Recommend a few work-friendly coffee shops where taking video calls wouldn't be awkward.
\item \textit{nomad\_4}: Where do remote workers in \textsf{\{area\}} usually go to work? I want somewhere with reliable internet where nursing one coffee for three hours is acceptable.
\item \textit{nomad\_5}: My coworking pass ran out and I need a cafe in \textsf{\{area\}} to work from tomorrow. Priorities: fast wifi, comfortable seating, not too crowded in the morning. Any suggestions?
\item \textit{nomad\_6}: Looking for a quiet spot with good coffee in \textsf{\{area\}} to write my report today --- somewhere I can plug in my laptop and focus. What do you recommend?
\end{itemize}

\subsection*{Couple --- date night}
\begin{itemize}\itemsep2pt
\item \textit{date\_1}: I just flew into \textsf{\{area\}} with my girlfriend and I want to take her somewhere special for dinner tonight. Something romantic with a nice atmosphere --- where should we go?
\item \textit{date\_2}: Most romantic restaurants in \textsf{\{area\}} for a date? Ideally cozy, good food, not a tourist trap.
\item \textit{date\_3}: It's our anniversary and we're in \textsf{\{area\}}. I'm looking for a memorable dinner spot --- beautiful setting, great food, doesn't have to be cheap. Recommendations?
\item \textit{date\_4}: Where can I take someone on a first date in \textsf{\{area\}}? Thinking a place with good drinks and a relaxed vibe where we can actually talk.
\item \textit{date\_5}: My partner and I (both around 30) want a nice sunset dinner in \textsf{\{area\}} tonight. What restaurant would you book?
\item \textit{date\_6}: Planning a surprise date night in \textsf{\{area\}} --- a restaurant with candles-and-good-wine energy rather than a party crowd. Where do locals actually go for this?
\end{itemize}

\subsection*{Professional --- business meeting}
\begin{itemize}\itemsep2pt
\item \textit{biz\_1}: I need to host a business lunch in \textsf{\{area\}} next week with two potential partners. Somewhere quiet and professional where we can talk without shouting. What do you suggest?
\item \textit{biz\_2}: Best restaurants in \textsf{\{area\}} for a business meeting? Needs to feel put-together, with tables spaced far enough apart for a private conversation.
\item \textit{biz\_3}: I'm meeting an investor for coffee in \textsf{\{area\}} tomorrow morning. Recommend a calm, presentable cafe that won't be blasting music at 9am.
\item \textit{biz\_4}: Where would you take a client to dinner in \textsf{\{area\}}? I want reliable service and food that impresses without being flashy.
\item \textit{biz\_5}: Our small team (4 people) is doing an offsite in \textsf{\{area\}} and we need a spot for a working lunch --- decent food, ok to sit for two hours with notebooks out. Ideas?
\item \textit{biz\_6}: Quiet place in \textsf{\{area\}} suitable for a serious one-on-one conversation over lunch --- I'm discussing a contract and need to hear the other person. Where should we meet?
\end{itemize}

\subsection*{Backpacker --- cheap and tasty}
\begin{itemize}\itemsep2pt
\item \textit{budget\_1}: I'm backpacking through \textsf{\{area\}} on a tight budget. Where can I eat tasty local food really cheap?
\item \textit{budget\_2}: Best cheap eats in \textsf{\{area\}}? I'd rather pay a few dollars for amazing local food than sit in an overpriced tourist place.
\item \textit{budget\_3}: I'm 24, traveling long-term, and my food budget in \textsf{\{area\}} is small. Which spots give the best value for money?
\item \textit{budget\_4}: Where do locals actually eat in \textsf{\{area\}}? Looking for authentic, filling meals that don't cost much.
\item \textit{budget\_5}: Recommend budget-friendly places in \textsf{\{area\}} for lunch --- big portions, low prices, and safe for a foreigner's stomach.
\item \textit{budget\_6}: My hostel is in \textsf{\{area\}} and I need a go-to place for cheap breakfasts and dinners nearby. What are the best inexpensive options?
\end{itemize}

\subsection*{Family with kids}
\begin{itemize}\itemsep2pt
\item \textit{family\_1}: We're a family with two kids (3 and 7) staying in \textsf{\{area\}}. Where can we have a relaxed dinner where the children won't drive other guests crazy?
\item \textit{family\_2}: Best kid-friendly restaurants in \textsf{\{area\}}? Bonus points for a play area or garden where they can run around while we finish eating.
\item \textit{family\_3}: Traveling with a toddler in \textsf{\{area\}} --- which cafes have high chairs and enough space for a stroller?
\item \textit{family\_4}: Sunday lunch spot in \textsf{\{area\}} for a family of five, grandparents included. Needs variety on the menu so everyone finds something. Suggestions?
\item \textit{family\_5}: My kids are picky eaters. Where in \textsf{\{area\}} can we get good food for adults and simple dishes (pasta, pancakes, smoothies) for children?
\item \textit{family\_6}: We want a casual family breakfast place in \textsf{\{area\}} --- quick service before the kids melt down, and decent coffee for the parents. Where should we go?
\end{itemize}

\subsection*{Vegan / dietary needs}
\begin{itemize}\itemsep2pt
\item \textit{vegan\_1}: I'm vegan and just moved to \textsf{\{area\}} for the season. Which restaurants have genuinely good plant-based food, not just one sad salad on the menu?
\item \textit{vegan\_2}: Best vegan or vegetarian restaurants in \textsf{\{area\}}? I eat out daily so I need places with real variety.
\item \textit{vegan\_3}: I'm gluten-free and my partner is vegetarian. Where can we both eat well in \textsf{\{area\}} without interrogating the waiter about every dish?
\item \textit{vegan\_4}: Healthy food spots in \textsf{\{area\}} --- smoothie bowls, salads, grilled fish, that kind of thing. Where do the health-conscious crowd actually eat?
\item \textit{vegan\_5}: Recommend a plant-based breakfast place in \textsf{\{area\}}. I care about ingredient quality and good coffee with oat milk.
\item \textit{vegan\_6}: I have a dairy allergy (not a preference --- an allergy). Which restaurants in \textsf{\{area\}} are careful and clearly label what's in their food?
\end{itemize}

\subsection*{Specialty coffee seeker}
\begin{itemize}\itemsep2pt
\item \textit{coffee\_1}: I'm serious about coffee. Where in \textsf{\{area\}} can I get properly made specialty coffee --- good beans, baristas who know what they're doing?
\item \textit{coffee\_2}: Best specialty coffee shops in \textsf{\{area\}}? I'm looking for local roasters and single-origin pour-overs, not just a pretty interior.
\item \textit{coffee\_3}: I'm a barista on holiday in \textsf{\{area\}}. Which cafes here would impress someone who works in specialty coffee?
\item \textit{coffee\_4}: Where can I find the best flat white in \textsf{\{area\}}? Milk texture matters to me more than latte art.
\item \textit{coffee\_5}: Recommend coffee shops in \textsf{\{area\}} that roast their own beans --- I want to buy a bag to take home too.
\item \textit{coffee\_6}: My morning ritual is a long black and a pastry. Which cafe in \textsf{\{area\}} should become my daily spot for the next two weeks?
\end{itemize}

\subsection*{Late-night group}
\begin{itemize}\itemsep2pt
\item \textit{night\_1}: It's 11pm, I'm in \textsf{\{area\}} with four friends and we're hungry. What's actually still open and good right now?
\item \textit{night\_2}: Best late-night food in \textsf{\{area\}}? We usually finish at the bar around midnight and need something better than instant noodles.
\item \textit{night\_3}: Where should a group of six go in \textsf{\{area\}} for dinner and drinks that flows into a fun night? We don't want to change venues three times.
\item \textit{night\_4}: Bars in \textsf{\{area\}} with genuinely good food, not just fries? We want cocktails and a proper meal in one place.
\item \textit{night\_5}: I land in \textsf{\{area\}} at 10pm tonight. Where can I get a solid meal near midnight --- somewhere reliable, not a gamble?
\item \textit{night\_6}: Birthday night out in \textsf{\{area\}} for ten people --- a place that takes big-group reservations, with music but where the kitchen stays open late. Suggestions?
\end{itemize}

\end{document}